\documentclass[fleqn,pra,twocolumn]{revtex4}

\usepackage{amsmath,graphicx}
\usepackage{dcolumn}
\usepackage{verbatim}

\begin{document}

\title{Fundamental transitions and ionization energies of the hydrogen molecular ions at the few ppt level.}

\author{Vladimir I. Korobov}
\affiliation{Bogoliubov Laboratory of Theoretical Physics, Joint Institute
for Nuclear Research, Dubna 141980, Russia}
\author{L.~Hilico}
\author{J.-Ph.~Karr}
\affiliation{Laboratoire Kastler Brossel, UPMC-Univ. Paris 6, ENS, CNRS, Coll\`ege de France\\
4 place Jussieu, F-75005 Paris, France}
\affiliation{Universit\'e d'Evry-Val d'Essonne, Boulevard Fran\c cois Mitterrand, F-91000 Evry, France}

\begin{abstract}
We calculate ionization energies and fundamental vibrational transitions for H$_2^+$, D$_2^+$, and HD$^+$ molecular ions. The NRQED expansion for the energy in terms of the fine structure constant $\alpha$ is used. Previous calculations of orders $m\alpha^6$ and $m\alpha^7$ are improved by including second-order contributions due to the vibrational motion of nuclei. Furthermore, we evaluate the largest corrections at the order $m\alpha^8$. That allows to reduce the fractional uncertainty to the level of $7.6\times10^{-12}$ for fundamental transitions and to $4.5\times10^{-12}$ for the ionization energies.
\end{abstract}

\maketitle

The hydrogen molecular ions (HMI) play an essential role in testing molecular quantum mechanics \cite{Carrington89,Leach95}. From the theoretical point of view the HMI is one of the simplest nonintegrable quantum system, which still allows very accurate numerical treatment. As was pointed out already some time ago \cite{Wing76}, and recently discussed more extensively \cite{Karr16}, if the theory would be sufficiently precise, the spectroscopy of HMI may be used for determining of the fundamental physical constants such as the proton-to-electron mass ratio. The ionization energy of HMI is also of high importance for the determination of ionization and dissociation energies of the hydrogen molecule from spectroscopic studies \cite{MerktH2,MerktD2,MerktHD} as well as for the determination of atomic masses of light nuclei \cite{Myers15,VanDyck15,codata14}.

On the experimental side there are many new projects started, which are now oriented towards Doppler-free spectroscopy with accuracy targeted to 1 ppt (one part per trillion) or better~\cite{Karr16,Tran13,Schiller14,Schiller17}. These perspectives bring strong motivation for theory.

The aim of this Letter is to improve the theoretical precision of spin-averaged energies and ro-vibrational transition frequencies in HMI. To this end we consider the largest QED contributions which had not been evaluated in our previous works \cite{KorobovPRA08,KorobovPRA14,KorobovPRL14}, namely, corrections to orders $m\alpha^6$ and $m\alpha^7$ due to vibrational motion of nuclei and the leading contributions to order $m\alpha^8$. As it was shown recently \cite{KorobovPRL16}, taking into account the vibrational motion of nuclei is essential for accurate theoretical description. It has allowed to resolve the longstanding discrepancy between theory and experiment in the hyperfine structure of H$_2^+$ ion. These new achievements reduce the relative uncertainty in the fundamental vibrational transitions of HMI to the level of $7.6\times10^{-12}$ and allow to obtain the most precise theoretical values for the ionization energies of H$_2^+$, D$_2^+$, and HD$^+$ molecular ions. In conclusion we discuss how these new results may have impact on fundamental physical constants such as the Rydberg constant, proton-to-electron mass ratio, and proton charge radius.

We use atomic units throughout this paper.

\vspace{3mm}
The terms of $m\alpha^6$ and higher orders are calculated in the adiabatic approximation. For this purpose we use the Born-Oppenheimer formalism. In this approach the states of the molecule are taken in the form
\begin{equation}\label{BO}
\Psi^{\rm BO} = \phi_{\rm el}(\mathbf{r};R)\chi_{\rm BO}^{}(R)
\end{equation}
The electronic wave function obeys the clamped nuclei Schr\"odinger equation for a bound electron
\begin{equation}\label{BO_el}
\left[H_{\rm el}-\mathcal{E}_{\rm el}(R)\right]\phi_{\rm el}=0,
\end{equation}
where $H_{el} = p^2/(2m) + V +Z_1Z_2/R$ is the electronic Hamiltonian, $V = -Z_1/r_1 - Z_2/r_2$, where $Z_1$ and $Z_2$ are the charges of the nuclei and $r_1$, $r_2$ are the distances from the electron to nuclei 1 and 2, respectively.
The wave function $\chi_{\rm BO}^{}(R)$ describes the relative nuclear motion, and is a solution of
\begin{equation}\label{radial}
\left(H_{vb}\!-\!E_0\right)\chi_{\rm BO}^{}
 = \left[-\frac{\nabla_R^2}{2\mu_n}\!+\!\mathcal{E}_{\rm el}(R)\!-\!E_0\right]\chi_{\rm BO}^{}
 = 0,
\end{equation}
where $\mu_n$ is the reduced mass of the nuclei.

Relativistic corrections of order $m\alpha^6$ to the energy of a bound electron in the two-center problem are
determined \cite{Pachucki97,KorobovJPB07} by the effective Hamiltonian:
\begin{equation}\label{EffH}
\begin{array}{@{}l}
\displaystyle
H^{(6)} = \frac{p^6}{16m^5}
     +\frac{[\boldsymbol{\nabla}V]^2}{8m^3}
     -\frac{3\pi}{16m^4}
        \Bigl\{
           p^2\rho+\rho\,p^2
        \Bigr\}
\\[2mm]\displaystyle\hspace{10mm}
     +\frac{5}{128m^4}\left(p^4V+Vp^4\right)
     -\frac{5}{64m^4}\left(p^2Vp^2\right),
\end{array}
\end{equation}
and the second order contribution of the Breit-Pauli Hamiltonian,
\begin{equation}\label{HB2}
\Delta E_B = \left\langle H_B Q (E_{\rm el}-H_{\rm el})^{-1} Q H_B \right\rangle.
\end{equation}
Here $\rho=\boldsymbol{\nabla}^2V/(4\pi)$, $Q$ is a projection operator onto a subspace orthogonal to $\phi_{\rm el}$ from Eq.~(\ref{BO_el}). $H_B$ is the Breit--Pauli relativistic correction for a bound electron:
\begin{equation}
H_B = -\frac{p^4}{8m^3}
   + \frac{\pi\rho}{2m^2} + H_B^{so},
\end{equation}
$H_B^{so}$ is the electron spin-orbit contribution (see details in \cite{KorobovJPB07}).
Both terms are divergent but their sum is finite
\begin{equation}\label{a6el}
\begin{array}{@{}l}
\mathcal{E}_{rc}^{(6)}(R) = 
   \alpha^4\Bigl[
      \Delta E_B(R)
      +\left\langle H^{(6)} \right\rangle(R)
   \Bigr].
\end{array}
\end{equation}

The leading contribution was obtained in \cite{KorobovPRA08} by averaging this effective potential over $R$:
\begin{subequations}
\begin{equation}\label{E6}
\Delta E^{(6)}_{el} =
   \left\langle\chi_{\rm BO}|\mathcal{E}_{rc}^{(6)}(R)|\chi_{\rm BO}\right\rangle.
\end{equation}

The next step is to consider the three-body correction to the energy $E_0$ of a molecular state, which we derive within the framework of the adiabatic approximation defined by Eqs.~(\ref{BO})-(\ref{radial}). This correction stems from the insertion of the Breit-Pauli effective potential $\mathcal{E}_B(R) = \alpha^2 \langle H_B \rangle$ into Eq.~(3) and in the order $m\alpha^6$ is expressed by
\begin{equation}\label{E6v}
\begin{array}{@{}l}\displaystyle
\Delta E^{(6)}_{vb} =
\\[2mm]\displaystyle\hspace{5mm}
   \left\langle\chi_{\rm BO}|\mathcal{E}_B(R)Q' (E_0\!-\!H_{vb})^{-1}Q'\mathcal{E}_B(R)|\chi_{\rm BO}\right\rangle,
\end{array}
\end{equation}
\end{subequations}
$Q'$ is a projection operator onto a subspace orthogonal to $\chi_{\rm BO}^{}(R)$.

Obviously, instead of the Born-Oppenheimer solution $\chi_{\rm BO}^{}(R)$ one may use the adiabatic solution $\chi_{\rm ad}(R)$, which includes as well the adiabatic corrections (see for definitions Ref.~\cite{Wolniewicz80}, or a review by Carrington {\em et al.} \cite{Carrington89}).

A complete set of the contributions at order $m\alpha^6$ is presented in Table \ref{tab:a6}. We include here as well the relativistic recoil contribution at order $m(Z\alpha)^6(m/M)$ \cite{PachuckiPRL97} and the radiative recoil contribution \cite{Pachucki95,Melnikov01}. In the former case the part, which depends on the state wave function, was evaluated with the help of LCAO approximation and its value had been used as an error bar for the recoil term.

The total contribution to the one-loop self energy correction {\em at order $m\alpha^7$} similarly should be written
\begin{equation}\label{a7se}
\hspace*{-2mm}
\begin{array}{@{}l}\displaystyle
\Delta E^{(7)}_{el} =
   \left\langle\chi_{\rm ad}|\mathcal{E}_{1loop-{\rm SE}}^{(7)}(R)|\chi_{\rm ad}\right\rangle,
\\[2mm]\displaystyle
\Delta E^{(7)}_{vb} =
   \left\langle\chi_{\rm ad}|
      \mathcal{E}_B(R)Q' (E_0\!-\!H_{vb})^{-1}Q'\mathcal{E}^{(5)}_{\rm SE}(R)
   |\chi_{\rm ad}\right\rangle,
\end{array}
\hspace*{-2mm}
\end{equation}
where $\mathcal{E}_{1loop-{\rm SE}}^{(7)}(R)$ is an effective potential of the $m\alpha^7$-order correction (see Eq.~(11), in \cite{KorobovPRA14}), to the energy of the bound electron in the two-center problem, and
\begin{equation}\label{a5se}
\begin{array}{@{}l}\displaystyle
\mathcal{E}_{\rm SE}^{(5)}(R) = \alpha^3\frac{4}{3}
   \left[\ln{\frac{1}{\alpha^2}}-{\beta(R)}+\frac{5}{6}-\frac{3}{8}\right]
\\[3mm]\displaystyle\hspace{30mm}
   \times\left\langle
      Z_1\delta(\mathbf{r}_1)\!+\!Z_2\delta(\mathbf{r}_2)
   \right\rangle
\end{array}
\end{equation}
is the one-loop self-energy correction of order $m\alpha^5$. $\beta(R)$ is the nonrelativistic Bethe logarithm for the bound electron in the two-center problem, whose values as a function of $R$ may be found in the Supplemental Material to Ref.~\cite{Korobov13} or in \cite{Kolos}.

A similar separation between electronic and vibrational contributions also occurs for the one-loop vacuum polarization term, which was obtained in \cite{Karr17}.

\begin{table}[t]
\begin{center}
\caption{Summary of contributions at order $m\alpha^6$ to the fundamental transitions in $\mbox{H}_2^+$, $\mbox{D}_2^+$, and $\mbox{HD}^+$ (in kHz). 
The first four contributions are defined as written in Eq.~(1) of \cite{KorobovPRA08}, $\Delta E_{rel-el}$ is the electronic contribution from Eq.~(\ref{E6}), $\Delta E_{rel-vb}$ is the newly obtained vibrational contribution from Eq.~(\ref{E6v}). The last contribution is the sum of the recoil and radiative-recoil corrections of order $m\alpha^6 (m/M)$ (see text).}\label{tab:a6}
\begin{tabular}{l@{\hspace{5mm}}d@{\hspace{5mm}}d@{\hspace{5mm}}d}
\hline\hline
\vrule height 10.5pt width 0pt depth 3.5pt
 & \mbox{H}_2^+\hspace*{-3mm} & \mbox{D}_2^+\hspace*{-3mm} & \mbox{HD}^+\hspace*{-5mm} \\
\hline
\vrule height 10pt width 0pt
$\Delta E_{1loop-{\rm SE}}$&-1881.2 &  -1362.3   &  -1647.0   \\
$\Delta E_{anom}$      &     21.2   &     15.3   &     18.5   \\
$\Delta E_{\rm VP}$    &    -66.3   &    -48.0   &    -58.0   \\
$\Delta E_{2loop}$     &    -55.9   &    -40.5   &    -48.9   \\
$\Delta E_{rel-el}$    &    -15.1   &    -10.5   &    -13.0   \\
$\Delta E_{rel-vb}$    &     44.6   &     32.2   &     39.0   \\
$\Delta E_{recoil}$    &     0.75(3)&    0.27(1) &      0.49(2) \\
\hline
\vrule height 10pt width 0pt
$\Delta E_{tot}$& -1952.0(1) & -1413.4(1) & -1708.9(1) \\
\hline\hline
\end{tabular}
\end{center}
\vspace*{-3mm}
\end{table}

Contributions to order $m\alpha^7$ without the vibrational second-order term were considered in \cite{KorobovPRA14,KorobovPRL14}. Here we present final results which appear in Table \ref{tab:a7}. We have managed to significantly improve precision of the relativistic correction to the Bethe logarithm (see, for details, \cite{KorobovHI15}), which allowed to reduce the theoretical uncertainty in the one-loop self-energy by an order of magnitude.

\begin{table}[t]
\begin{center}
\caption{Summary of contributions at order $m\alpha^7$ to the fundamental transitions in $\mbox{H}_2^+$, $\mbox{D}_2^+$, and $\mbox{HD}^+$ (in kHz). The first two contributions are the one-loop self-energy and vacuum polarization corrections, which include the vibrational contribution (see text). The last three contributions are defined in Eqs. (23)--(25) of \cite{KorobovPRA14}.}\label{tab:a7}
\begin{tabular}{l@{\hspace{9mm}}d@{\hspace{9mm}}d@{\hspace{9mm}}d}
\hline\hline
\vrule height 10.5pt width 0pt depth 3.5pt
 & \mbox{H}_2^+\hspace*{-3mm} & \mbox{D}_2^+\hspace*{-3mm} & \mbox{HD}^+\hspace*{-5mm} \\
\hline
\vrule height 10pt width 0pt
$\Delta E_{1loop-{\rm SE}}$ & 109.0(1) &  78.8(1)  &  95.4(1)  \\
$\Delta E_{\rm VP}$      &      2.8    &   2.0     &   2.4     \\
$\Delta E_{\rm WK}$      &     -0.08   &  -0.06    &  -0.07    \\
$\Delta E_{2loop}$       &     10.1    &   7.3     &   8.9     \\
$\Delta E_{3loop}$       &     -0.06   &  -0.05    &  -0.05    \\
\hline
\vrule height 10pt width 0pt
$\Delta E_{tot}$         &   121.8(1)  &  88.1(1)  & 106.4(1) \\
\hline\hline
\end{tabular}
\end{center}
\vspace*{-3mm}
\end{table}

Finally, we turn to the evaluation of {\em $m\alpha^8$-order corrections}.

\begin{table}[t]
\begin{center}
\caption{Summary of contributions at order $m\alpha^8$ to the fundamental transitions
in $\mbox{H}_2^+$, $\mbox{D}_2^+$, and $\mbox{HD}^+$ (in kHz).}\label{tab:a8}
\begin{tabular}{l@{\hspace{3mm}}d@{\hspace{3mm}}d@{\hspace{3mm}}d}
\hline\hline
\vrule height 10.5pt width 0pt depth 3.5pt
 & \mbox{H}_2^+\hspace*{-3mm} & \mbox{D}_2^+\hspace*{-3mm} & \mbox{HD}^+\hspace*{-5mm} \\
\hline
\vrule height 10pt width 0pt
$\Delta E_{2loop}$ &   -1.34(21) & -0.97(15) & -1.17(18)\\
$\Delta E_{1loop-{\rm SE}}$& -0.97(48) & -0.70(35) & -0.85(42)\\
$\Delta E_{\rm VP}$    &   -0.017    & -0.013    & -0.015   \\
\hline
\vrule height 10pt width 0pt
$\Delta E_{tot}$   &   -2.3(5)   & -1.7(4)   & -2.0(5) \\
\hline\hline
\end{tabular}
\end{center}
\end{table}

For hydrogen-like atoms, the two-loop correction at order $m\alpha^8$ may be written in the form
\begin{equation}
\begin{array}{@{}l}\displaystyle
E_{2loop}^{(8)} = \left(\frac{1}{\pi}\right)^2\!\frac{(Z\alpha)^6}{n^3}
   \Bigl[
      B_{63}L^3(Z\alpha)\!
      +\!B_{62}L^2(Z\alpha)\!
\\[3mm]\displaystyle\hspace{30mm}
      +B_{61}L(Z\alpha)+B_{60}
   \Bigr],
\end{array}
\end{equation}
where $L(Z\alpha)\equiv\ln(Z\alpha)^{-2}$. It is useful to recall the numerical values of the various terms for the ground state of the hydrogen atom \cite{Eides_book}:
\[
\Delta E(1S)\approx
   \frac{(Z\alpha)^6}{\pi^2}
   \left[{-282-62+476-61}\right].
\]
This shows that the third term (linear in $\ln(Z\alpha)^{-2}$) is the largest one, contrary to our intuition on hierarchy of the consecutive terms in the $Z\alpha$ expansion.

In case of a two-center system the corrections can still be written in the form of Eq. (11) (with $n\!=\!1$ and $Z_1\!=\!Z_2\!=\!Z$). The first three coefficients $B_{6k}$ can be obtained from the results of \cite{JCP05} as
\begin{equation}\label{b62}
\begin{array}{@{}l}\displaystyle
Z^6B_{63} = -\frac{8}{27}\,Z^3\pi\bigl\langle\delta(\mathbf{r}_1)\!+\!\delta(\mathbf{r}_2)\bigr\rangle,
\\[2.5mm]\displaystyle
Z^6B_{62} =
   \frac{1}{9}\left\langle
      \boldsymbol{\nabla}^2V\>Q(E_0-H)^{-1}Q\>\boldsymbol{\nabla}^2V
   \right\rangle_{\rm fin}
\\[1.5mm]\hspace{20mm}\displaystyle
   +\frac{1}{18}\left\langle
      \boldsymbol{\nabla}^4V
   \right\rangle_{\rm fin}
\\[2mm]\hspace{14mm}\displaystyle
   +\frac{16}{9}\left[\frac{31}{15}+2\ln{2}\right]\,Z^3\pi
      \bigl\langle\delta(\mathbf{r}_1)\!+\!\delta(\mathbf{r}_2)\bigr\rangle,
\end{array}
\hspace*{-5mm}
\end{equation}
and
\begin{equation}\label{b61}
\begin{array}{@{}l}\displaystyle
Z^6B_{61} =
   -2\Bigl[
      \frac{1}{9}\left\langle
         \boldsymbol{\nabla}^2V\>Q(E_0-H)^{-1}Q\>\boldsymbol{\nabla}^2V
      \right\rangle_{\rm fin}
\\[3mm]\displaystyle\hspace{12mm}
      +\frac{1}{18}\left\langle \boldsymbol{\nabla}^4V \right\rangle_{\rm fin}
   \Bigr]\ln{2}
   +\frac{4}{3}N(R)
\\[3mm]\displaystyle\hspace{12mm}
   +\frac{19}{135}\left\langle
      \boldsymbol{\nabla}^2V\>Q(E_0-H)^{-1}Q\>\boldsymbol{\nabla}^2V
   \right\rangle_{\rm fin}
\\[3mm]\displaystyle\hspace{12mm}
   +\frac{19}{270}\left\langle \boldsymbol{\nabla}^4V \right\rangle_{\rm fin}
   +\frac{1}{24}\left\langle
      2\mathrm{i}\sigma_{}^{ij}p^i\boldsymbol{\nabla}^2Vp^j
   \right\rangle
\\[4mm]\displaystyle\hspace{12mm}
   +\Bigl[
      \frac{48781}{64800}\!+\!\frac{2027\pi^2}{864}
      \!+\!\frac{56}{27}\ln{2}\!-\!\frac{2\pi^2}{3}\ln{2}\!
\\[3mm]\displaystyle\hspace{15mm}
      +\!8\ln^2{2}\!+\!\zeta(3)
   \Bigr]Z^3\pi\bigl\langle\delta(\mathbf{r}_2)\!+\!\delta(\mathbf{r}_2)\bigr\rangle.
\end{array}
\end{equation}

\begin{table*}[t]
\begin{center}
\caption{Fundamental transition frequencies $\nu_{01}$ for $\mbox{H}_2^+$, $\mbox{D}_2^+$, and $\mbox{HD}^+$ molecular ions (in kHz). CODATA14 recommended values of constants. The first error is the theoretical uncertainty, the second error is due to the uncertainty in mass ratios.}\label{ftE}
\begin{tabular}{l@{\hspace{16mm}}d@{\hspace{16mm}}d@{\hspace{16mm}}d}
\hline\hline
\vrule height 10.5pt width 0pt depth 3.5pt
 & \mbox{H}_2^+\hspace*{3mm} & \mbox{D}_2^+\hspace*{3mm} & \mbox{HD}^+\hspace*{2mm} \\
\hline
\vrule height 10pt width 0pt
$\nu_{nr}$ & 65\,687\,511\,047.0   & 47\,279\,387\,818.4   & 57\,349\,439\,952.4   \\
$\nu_{\alpha^2}$ &   1091\,040.5   &          795\,376.3   &          958\,151.7   \\
$\nu_{\alpha^3}$ &   -276\,545.1   &         -200\,278.0   &         -242\,126.3   \\
$\nu_{\alpha^4}$ &       -1952.0(1)&             -1413.4(1)&             -1708.9(1)\\
$\nu_{\alpha^5}$ &         121.8(1)&                88.1(1)&               106.4(1)\\
$\nu_{\alpha^6}$ &          -2.3(5)&                -1.7(4)&                -2.0(5)\\
\hline
\vrule height 10pt width 0pt
$\nu_{tot}$& 65\,688\,323\,710.1(5)(29) & 47\,279\,981\,589.8(4)(8) & 57\,350\,154\,373.4(5)(17) \\
\hline\hline
\end{tabular}
\end{center}
\vspace*{-3mm}
\end{table*}

\begin{table*}[t]
\begin{center}
\caption{Ionization energies $E_I$ for $\mbox{H}_2^+$, $\mbox{D}_2^+$, and $\mbox{HD}^+$ molecular ions (in cm$^{-1}$). CODATA14 recommended values of constants. The error is the theoretical uncertainty. The error due to the uncertainty in mass ratio is below $10^{-7}$ cm$^{-1}$.} \label{IE}
\begin{tabular}{l@{\hspace{15mm}}d@{\hspace{15mm}}d@{\hspace{15mm}}d}
\hline\hline
\vrule height 10.5pt width 0pt depth 3.5pt
 & \mbox{H}_2^+\hspace*{-3mm} & \mbox{D}_2^+\hspace*{-3mm} & \mbox{HD}^+\hspace*{-5mm} \\
\hline
\vrule height 10pt width 0pt
$E_{I,nr}$ & 131\,056.875\,7465  & 131\,418.947\,7041  & 131\,223.436\,2578   \\
$E_{I,\alpha^2}$ &   1.599\,4995 &        1.604\,8306  &        1.601\,9142   \\
$E_{I,\alpha^3}$ &  -0.350\,9300 &       -0.352\,5527  &       -0.351\,6791   \\
$E_{I,\alpha^4}$ &  -0.002\,4774(1)&     -0.002\,4897(1)&      -0.002\,4831(1)\\
$E_{I,\alpha^5}$ &   0.000\,1569(1)&      0.000\,1576(1)&       0.000\,1409(1)\\
$E_{I,\alpha^6}$ &  -0.000\,0021(6)&     -0.000\,0021(6)&      -0.000\,0021(6)\\
\hline
\vrule height 10pt width 0pt
$E_{I,tot}$& 131\,058.121\,9937(6) & 131\,420.197\,6480(6) & 131\,224.684\,1650(6) \\
\hline\hline
\end{tabular}
\end{center}
\vspace*{-3mm}
\end{table*}

Among the terms presented in Eqs.~(\ref{b62}), (\ref{b61}) all the distributions were determined and calculated in \cite{KorobovPRA14} except $N$, which is defined in Eq. (4.21.a) of \cite{JCP05}. On the other hand the expression for $N$ is similar to the one of Eq.~(10) in Ref.~\cite{Korobov13}. Using the same technique, which has been used for calculations of the relativistic Bethe logarithm we were able to get $N$ for the hydrogen atom ground state:
\[
N(1S) = 17.8556720362(1),
\]
which is in a good agreement with the value given in \cite{JCP05} and even adds two more significant digits. Having validated our approach, we then did calculations of the $N(R)$ "effective" potential for the two-center problem. Putting it into Eq.~(\ref{b61}) and then averaging over $R$ we get for the {\em ionization energy} of H$_2^+$ (in kHz)
\[
\begin{array}{@{}l}\displaystyle
\Delta E^{(8)}_{2loop} = \alpha^6\left[
      B_{63}L^3(\alpha)\!+\!B_{62}L^2(\alpha)\!+\!B_{61}L(\alpha)\!+\!B_{60}
   \right]
\\[3mm]\displaystyle\hspace{15mm}
\approx 37.0-17.3-52.9+7.8.
\end{array}
\]
The last term in the second line has been evaluated in the LCAO approximation using the atomic hydrogen ground state value for $B_{60}$. We take the error bar on the two-loop contribution as equal to this approximate value of nonlogarithmic term. In our previous studies we used the same kind of estimates for the uncertainty resulted from the yet uncalculated terms and further improvements of the theory showed the good relevance of this approach.

Similarly, for the {\em fundamental transition} $(L\!=\!0,v\!=\!0)\to(L'\!=\!0,v'\!=\!1)$ (in kHz)
\[
\begin{array}{@{}l}\displaystyle
\Delta \nu^{(8)}_{2loop} = \alpha^6\left[
      B_{63}^{\nu}L^3(\alpha)\!+\!B_{62}^{\nu}L^2(\alpha)\!+\!B_{61}^{\nu}L(\alpha)\!+\!B_{60}^{\nu}
   \right]
\\[3mm]\displaystyle\hspace{15mm}
\approx 0.97-1.68-0.84+0.21,
\end{array}
\]
and for the uncertainty we take $u_r(E_{2loop})=0.21$ kHz.

The other significant contribution at the $m\alpha^8$ order is the one-loop self-energy,
\begin{equation}
E_{1loop}^{(8)} = \frac{\alpha^6}{\pi n^3}\,Z^7
   \left[
      A_{71}\ln(Z\alpha)^{-2}\!+\!A_{70}
   \right],
\end{equation}
where the leading term has analytic result \cite{Eides_book,Karshenboim97}: $A_{71}(nS) = \pi\left[139/64-\ln{2}\right]$.
For the hydrogen atom the nonlogarithmic contribution $A_{70}$ of order $m\alpha(Z\alpha)^7$ has never been calculated directly. By extrapolation of the $G_{se}(1S,Z\alpha)$ function \cite{A70} with the expansion over $Z\alpha$ (see Eq.~(5.1) from \cite{A70}) one may get $A_{70}=44.4$. Similarly to the two-loop corrections above, we take the nonlogarithmic term as estimate of the theoretical uncertainty..

The second order contributions due to vibrational motion, both from one- and two-loop diagrams, were evaluated as well. The total frequency shift is about 100 Hz and may be neglected for the time being.

\vspace{3mm}
The main results of our work, frequencies for the fundamental transitions $(L=0,v=0)\to(0,1)$ and ionization energies of the HMI, are presented in Tables \ref{ftE} and \ref{IE}, respectively. To get precision data for the relativistic corrections of order $m\alpha^4$ we have used the expectation values of the Breit-Pauli operators, which were obtained in \cite{YanH2+,YanHD+,YanD2+} with 15 or even more significant digits. As it may be extracted from the Tables the new theoretical relative uncertainty for the fundamental transition frequency is $u_r(\nu(\mbox{H}_2^+))=0.5/(66.\!\times\!10^{9})\approx 7.6\times10^{-12}$, and accordingly for the ionization energy one gets $u_r(E_I)=4.5\times10^{-12}$. The CODATA14 uncertainty of the Rydberg constant is $u_r(R_{\infty})=5.9\times10^{-12}$. Since this constant enters in the data of the Tables as a multiplier, an uncertainty in the energies due to the uncertainty in the Rydberg constant can be easily evaluated and is not shown.

These results have direct impact on the potential determination of the fundamental constants. For example, the theoretical uncertainty on the fundamental transition in H$_2^+$ sets the following limit on the achievable precision of the proton-to-electron mass ratio ($\mu_p=m_p/m_e$) to
\begin{equation}\label{mass}
\Delta\mu_p/\mu_p=1.5\times10^{-11}.
\end{equation}
This uncertainty is smaller by a factor of 6 with respect to the present CODATA, $u_r(\mu_p) = 9.5\times10^{-11}$ \cite{codata14}, which is currently limited by uncertainty on the proton's atomic mass. The electron's atomic mass has been recently improved ($u_r(A_r(e))=3.1\times10^{-11}$) by a high-precision measurement of the $g$-factor of a bound electron in a $^{12}$C$^{5+}$ ion \cite{Sturm14}. In terms of ultimate accuracy limits, the $1.5\times10^{-11}$ theoretical uncertainty that we have achieved for HMI spectroscopy is comparable to the current theoretical uncertainty of $1.3\times10^{-11}$ on the $g$-factor of $^{12}$C$^{5+}$ \cite{codata14,Pachucki05}.

The \emph{proton rms charge radius} ($r_p$) uncertainty as determined in the CODATA14 adjustment has a much smaller contribution $\sim\!5\times10^{-12}$  to the uncertainty in the fundamental transitions. However, replacing the CODATA value of $r_p$ with that obtained from muonic hydrogen spectroscopy \cite{Pohl10,Antognini13} leads to a 3 kHz blueshift of the transition, i.e.\ a relative shift of $5\times 10^{-11}$. If we assume that the \emph{muonic hydrogen adjusted} Rydberg constant should be used as proper constant when using the \emph{muonic hydrogen} proton radius \cite{Pohl13}, then we get a global shift of 1.1 kHz, which is still feasible for detection (see also the more detailed discussion in \cite{Karr16}).

Finally, since the fundamental transitions have potentiality to be used for adjustment of the fundamental constants we present here in explicit form the frequency dependence of transition lines on the masses and on the proton and deuteron charge radii:
\begin{subequations}
\begin{equation}\label{energy}
\hspace*{-2mm}
\begin{array}{@{}l}\displaystyle
\nu(\mbox{H}_2^+) = \nu_0(\mbox{H}_2^+)
   +{\textstyle\frac{\Delta R_\infty}{R_\infty}}\,\nu_0(\mbox{H}_2^+)
   +2(R_\infty c)\,\times
\\[2mm]\displaystyle\hspace{5mm}
   \Bigl[
      -2.55528\cdot10^{-6}\Delta \mu_p
      -8.117\cdot10^{-12}\Delta r_p
   \Bigr],
\end{array}
\end{equation}
where $\Delta R_\infty=R_\infty\!-\!R_{\infty,0}$, $\Delta \mu_p = \mu_p\!-\!\mu_{p,0}$ and $\Delta r_p = r_p^2-r_{p,0}^2$, here the subscript $0$ stands for the CODATA14 value, and $\nu_0$ is the transition frequency presented in Table \ref{ftE}, which was calculated with the CODATA14 values of the constants.
\begin{equation}\label{energy:b}
\hspace*{-2mm}
\begin{array}{@{}l}\displaystyle
\nu(\mbox{D}_2^+) = \nu_0(\mbox{D}_2^+)
   +{\textstyle\frac{\Delta R_\infty}{R_\infty}}\,\nu_0(\mbox{D}_2^+)
   +2(R_\infty c)\,\times
\\[2mm]\displaystyle\hspace{5mm}
   \Bigl[
      -9.37686\cdot10^{-7}\Delta \mu_d
      -5.877\cdot10^{-12}\Delta r_d
   \Bigr],
\end{array}
\end{equation}
here $\Delta\mu_d = \mu_d\!-\!\mu_{d,0}$ and $\Delta r_d = r_d^2-r_{d,0}^2$,
\begin{equation}\label{energy:c}
\hspace*{-3mm}
\begin{array}{@{}l}\displaystyle
\nu(\mbox{HD}^+) = \nu_0(\mbox{HD}^+)
   +{\textstyle\frac{\Delta R_\infty}{R_\infty}}\,\nu_0(\mbox{HD}^+)
   +2(R_\infty c)\,\times
\\[2mm]\displaystyle\hspace{4mm}
   \Bigl[
      -\!1.49998\!\cdot\!10^{-6}\Delta \mu_p
      \!-\!3.75470\!\cdot\!10^{-7}\Delta \mu_d
\\[2mm]\displaystyle\hspace{8mm}
      -3.555\cdot10^{-12}\Delta r_p
      -3.550\cdot10^{-12}\Delta r_d
   \Bigr].
\end{array}
\hspace*{-3mm}
\end{equation}
\end{subequations}
In the last equation $\Delta r_d$ may be in principle eliminated since the measured H-D isotope shift of the 1$S$-2$S$ transition \cite{Parthey10} determines the deuteron-proton charge radius difference \cite{Jentschura11,codata14}
\[
r_d^2-r_p^2 = 3.81948(37) \mbox{ fm}^2
\]
with much smaller error than the CODATA14 uncertainties for $r_p$ and $r_d$.

In summary, we have considered several new contributions to the binding energies of HMI, which result in an essential improvement of the theoretical uncertainty both for the ionization energies and for the transition frequencies of the HMI. This level of precision allows to use the HMI spectroscopy as an alternative way for determination of the fundamental physical constants.

\emph{Acknowledgements.} This work was supported by UPMC, which is gratefully acknowledged. J.-Ph. Karr acknowledges support as a fellow of the Institut Universitaire de France. V.I.K. acknowledges support of the Russian Foundation for Basic Research under Grant No.~15-02-01906-a. Authors also very much appreciate valuable discussions with K.~Pachucki.

\end{document}